\documentclass[preprint,authoryear,12pt]{elsarticle}

\usepackage{enumerate}
\usepackage{graphicx}
\usepackage{amsmath}
\usepackage{epstopdf}
\usepackage{url}
\journal{Astronomy and Computing}

\begin{document}

\begin{frontmatter}

\title{JSPAM: A restricted three-body code for simulating interacting galaxies}

\author[jw,ah]{John F. Wallin}
\author[ah]{Anthony J. Holincheck}
\author[ah]{Allen Harvey}
\address[jw]{Department of Physics and Astronomy \& Center for Computational Science, Middle Tennessee State University}
\address[ah]{School of Physics, Astronomy, and Computational Science, George Mason University}

\begin{abstract}
Restricted three-body codes have a proven ability to recreate much of the disturbed morphology of actual interacting galaxies.  As more sophisticated n-body models were developed and computer speed increased, restricted three-body codes fell out of favor.  However, their supporting role for performing wide searches of parameter space when fitting orbits to real systems demonstrates a continuing need for their use.  Here we present the model and algorithm used in the JSPAM code\footnote{\url{http://ascl.net/1511.002}}.   A precursor of this code was originally described in 1990, and was called SPAM.   We have recently updated the software with an alternate potential and a treatment of dynamical friction to more closely mimic the results from n-body tree codes.  The code is released publicly for use under the terms of the Academic Free License (``AFL'') v. 3.0 and has been added to the Astrophysics Source Code Library.
\end{abstract}


\end{frontmatter}

\section{Introduction}
Soon after galaxies beyond the Milky Way were first observed, it was noticed that some of them appeared to be in close proximity to one another.  Many of these close pairs and groups contained disturbed morphologies, the cause of which was debated for much of the middle part of the twentieth century.  Speculation as to the source of the peculiar features included gravitational and magnetic interactions.  One of the first papers to demonstrate that tidal distortions caused by gravity could produce the thin tails and bridges seen in interacting galaxies was \cite{pfleiderer}. The researchers used a simple model for the particles of the primary galaxy in the interaction. It was treated as a point mass and was surrounded by a disk of massless test particles. The secondary galaxy was represented by a point mass with no resolved disk. The particles in the primary disk were influenced by the gravity of both point masses, but not each other.  This allowed for a relatively easy set of calculations to determine the forces felt by each particle at each time step in a simulation.  

Later researchers such as \cite{tt72} would add a disk of test particles around the secondary galaxy and attempt to simulate models of major mergers.  The authors of that work were able to produce credible models of four well-known pairs of interacting galaxies. This seemed to answer conclusively that it was possible to recreate the general morphology of interacting systems, especially the tails and bridges, with gravitational tidal disturbances. These models are referred to by several names, but they will be called restricted three-body methods in this work. The results from these models showed that the more dramatic features of interacting galaxies are produced by slow, close passages along parabolic or elliptical orbits. Faster passages usually result in smaller features being generated. 

This paper announces the public release of a specific restricted three-body software implementation.  The precursor to the JSPAM code presented here was first presented in \cite{wallin90} as SPAM\footnote{SPAM is the Stellar Particle Animation Module.}.  It has been used since then for parameter studies \citep{wallin92} and to model specific systems \citep{smith92}, \citep{wallin94}, \citep{smith94}, \citep{smith10}.  The ``J'' in JSPAM indicates the software was ported to the Java\textsuperscript{TM} language.  This paper presents details of significant enhancements to all language versions of the code.  The code has been added to the Astrophysics Source Code Library \citep{ascl2015}.

Though more sophisticated n-body models with gas dynamics and stellar feedback have been developed, restricted three-body codes remain useful.  They can generate a rough match for a wide range of morphologies.  Finding this match by manually varying the orbit parameters can be a time consuming process, especially when running full n-body codes.  Automating the search by connecting a restricted three-body code to an optimization routine is a much quicker approach.  Follow-up simulations can then be performed using self consistent, n-body codes.  In the last decade, several researchers have attempted this approach, see \cite{smith10} for a brief summary of methods.  The most common optimization approach is to use a genetic algorithm or similar evolutionary code to run thousands of models in order to converge on best-fit parameters.  This multi-method modeling technique benefits from the existence of a fast, restricted three-body code.  We have implemented our code in multiple languages to assist with integration into larger systems.  For example, the FORTRAN code can be compiled as a library and linked directly to other FORTRAN and C/C++ codes.  The Java\textsuperscript{TM} version was integrated with a Citizen Science project called {\it Galaxy Zoo~: Mergers} that used an Applet running in the browser of each of the volunteers that contributed \citep{holinAAS}.

\section{Methods}

\subsection{Gravity in Restricted Three-Body Problem}

The restricted three-body model has been used to model interacting galaxies since the
early 1970's.   The seminal models by \cite{tt72} using this method first
demonstrated that observed tidal features could be reproduced using simple gravitational
interactions without the need to invoke more complicated physical processes.

Since the late 1980's, hierarchical tree codes have been commonly used to model interacting galaxies.
By reducing the n-body problem's computational complexity from $\sim O(n^2)$ to $\sim O(n \log n)$, 
problems that would be intractable have become possible.   Of course,
Moore's Law has also played a critical role in the evolution of models.
The inclusion of other physics beyond gravity, including gas dynamics, star formation and
feedback processes, and even radiative transfer have made these models commonly
used to understand processes in interaction and cosmology \citep{robertson}.

In many cases, it helps to have a detailed numerical model of a particular interacting galaxy system.   Being able to match dynamical time to stellar population, for example, can bring a better understanding of the evolutionary processes. Even with these advances, modeling an observed interacting galaxy is a time consuming process.  For self-consistent tree code runs, you need large numbers of particles to represent the halo, disk and bulge of each galaxy. Even a modest ($< 100k$ particle) model can take hours or even days on a workstation to run a single model. In general, dozens to even thousands of models are needed to match a single interacting galaxy system.  Prototyping systems such as IDENTIKIT \citep{barnes2009}, have been created to help make this process easier.

Restricted three-body models have computational advantages over full n-body models.  
In these models, the gravitational field is represented with a static potential around two moving
masses.   Stars are represented by non-interacting test particles that are initially orbiting within
these potentials.   As the two potentials pass each other, the tidal forces create the 
features seen in interacting systems.   With modern computers, a modest resolution 
run with thousands of particles can execute in just a few seconds using this method.    By using
the restricted three-body method, it is possible to easily produce hundreds of thousands
numerical models in a single day.    Although humans cannot easily look at this many models,
automatic optimization systems such as genetic algorithms can be used converge on 
the best fit model for particular systems, ({\it e.g.}, \cite{dejong}).

In many older codes, the gravitational potential in restricted three-body codes was represented
with a softened point mass.    Although the models produced with these codes had similar
tidal features to those seen in nature, using the results from the code to move to 
high resolution model with a self-consistent n-body code was difficult due to the large differences between a softened point mass potential and a self-consistent n-body code.   Dynamical friction during the interaction also causes the trajectories and resulting tidal features to differ.

In the code we are releasing, we include a new potential in addition to the softened point mass.
This potential is derived by using the initial conditions for a self-consistent n-body
model.  Details are described in section \ref{threecomp}.  We have also included
a parameterized version of dynamical friction in this code. Some researchers have included an analytic evaluation of more realistic potentials based on the Hernquist and NFW mass distributions.  Simulations run by \cite{dubinski} and \cite{petsch_nfw} also combined these potentials with dynamical friction following a similar prescription to this code.  Our key innovation is to make the simulations run almost as fast as the traditional softened point mass by sampling a ``unit'' mass distributed according to the Hernquist distribution and then use fast interpolation to look up central force acceleration.

With this code, it is possible to rapidly prototype galaxy collisions and then use
the results to create high resolution self-consistent  n-body models of interacting galaxies.

\subsection{Potentials}

The restricted three-body problem considered here is that of a set of massless test particles in a system with two massive bodies.  Each of the massive bodies represents the center of mass of a galaxy.  The massless test particles are distributed around the centers of mass to represent the disk material.  The test particles do not interact with each other.  The sum of the gravitational potentials from the massive particles determines the dynamics of the system.  

JSPAM provides two different potentials.  The corresponding acceleration for the softened point mass potential has an analytic expression.  The acceleration due to the new potential is made from sampling the potential from a full n-body model of a three-component disk galaxy containing a halo, disk, and bulge.

\subsubsection{Softened Point Mass Potential}

The softened point mass used is based on the original SPAM code described by \cite{wallin90}.   In this approach, we soften the acceleration of the massless particles using the form:

\begin{equation}
\vec{a} = - \frac{GM}{(r^2+\epsilon ^2)} \hat{r}
\label{spm-eqn}
\end{equation}

Giving a resulting potential of the form:
\begin{equation}
\phi(r) = \frac{GM}{\epsilon} \biggl[ \frac{\pi}{2} - \tan^{-1} \biggl(\frac{r}{\epsilon}\biggr) \biggr]
\end{equation}

 To model the more extended potential of a galaxy, simulations using the original SPAM code used a softening length typical about $0.3$ times the radius of the disk galaxy.    Later in section \ref{comparison}, we will show how this potential compares to the n-body interpolated potential and full n-body tree code runs.

\subsubsection{N-body Interpolated Potential}
\label{threecomp}

In order to create a more realistic potential, we have adapted a modified three-component
model of the mass distribution of a spiral galaxy.   A bulge, disk, and halo mass distribution
are initialized, and then a new potential function is derived.

For our code, we follow the methodology first described by \cite{hern3} to specify the mass distribution in each component.  The mass and the characteristic radii of each of the three components is set in the code.   

Following \cite{hern3} we define $r_c$ as a cutoff radius, $\gamma$ as a core radius, and $\alpha$ as a normalization constant.  For the Halo we use:
\begin{eqnarray}
q_{halo} = \frac{\gamma_{halo} } {r_c}
%\nonumber
\end{eqnarray}

\begin{eqnarray}
\alpha_{halo} = 
\frac{1}
{1 -  q_{halo}  \sqrt{\pi}\exp \left[q_{halo}^2 \right]  (1  - {\rm erf}  (q_{halo} ))}
%\nonumber
\end{eqnarray}

\begin{eqnarray}
M(r)_{halo}  = M_{halo}
\frac{  2  \alpha_{halo} }
{ \sqrt{\pi} }
 \int_0^{\frac{r}{r_c}}  \frac{ \exp(-x^2)} {x^2 + q_{halo}^2} x^2 dx 
%\nonumber
\end{eqnarray}

For the bulge, we use a Gaussian distribution of mass, rather than the specific formula in \cite{hern3}.  We define $h_{bulge}$ as a scale length and include normalization constants for the integral.

\begin{eqnarray}
M(r)_{bulge}  = 
  M_{bulge}{\frac{4}{\sqrt{\pi}}}
 \int_0^{\frac{r}{h_{bulge}}}   exp (-x^2) x^2 dx
%\nonumber
\end{eqnarray}

And for the disk, we define $h_{disk}$ as a scale length and make the approximation that the mass is distributed with

\begin{eqnarray}
M(r)_{disk} = \frac{ M_{disk}}{2}  \int_0^{\frac{r}{h_{disk}}} exp (-x) x^2 dx
%\nonumber
\end{eqnarray}

While this approximation captures the drop in mass within the plane of the disk, it does not capture the true three dimensional distribution of the mass.   For particles orbiting in the disk, the approximation works very well.   However, for particles outside of the disk's plane, the approximation is less valid, particularly for those particles in the inner region.   For most of the collisions we have modeled, this approximation seems to create models that are qualitatively similar to those from hierarchical tree codes.  In part this is due to the velocity profile for this potential being flatter than that for the softened point mass potential, discussed in Section \ref{veloc}.  The computational advantages of using a spherical mass distribution seem to outweigh the need for a fully realistic potential in these models.

Summing the mass of all the components, we can then calculate the potential
\begin{eqnarray}
M_{total}(r) = M_{disk}(r) + M_{halo}(r) + M_{bulge}(r)
%\nonumber
\end{eqnarray}

\begin{eqnarray}
\vec{a} = \frac{G M_{total} (r) }{r^2} \hat{r} 
\label{nbi-eqn}
\end{eqnarray}

For all of our simulations, the potential is initialized using a ``unit'' galaxy.  The relative masses for this initial potential are 5.8 for the halo, 1 for the disk, and 0.3333 for the bulge.  The mass-dependent values are normalized to 1 by dividing by the sum of all three components, 7.1333.  The scale length for the disk, $h_{disk}$, is set to 1.  The scale lengths for the halo, $r_c$ and $r_{halo}$, are set to 10, and $\gamma_{halo}$ is set to 1.  The scale length for the bulge, $h_{bulge}$, is set to 0.2.  This ``unit'' galaxy can then be scaled to match the mass and disk radius specified for any specific galaxy to be simulated.

\subsubsection{Velocity Profile}
\label{veloc}
\begin{figure}[htbp]
%\begin{minipage}{0.8\linewidth}
\begin{center}
%\plotone{figure1.eps}
\includegraphics[scale=0.7]{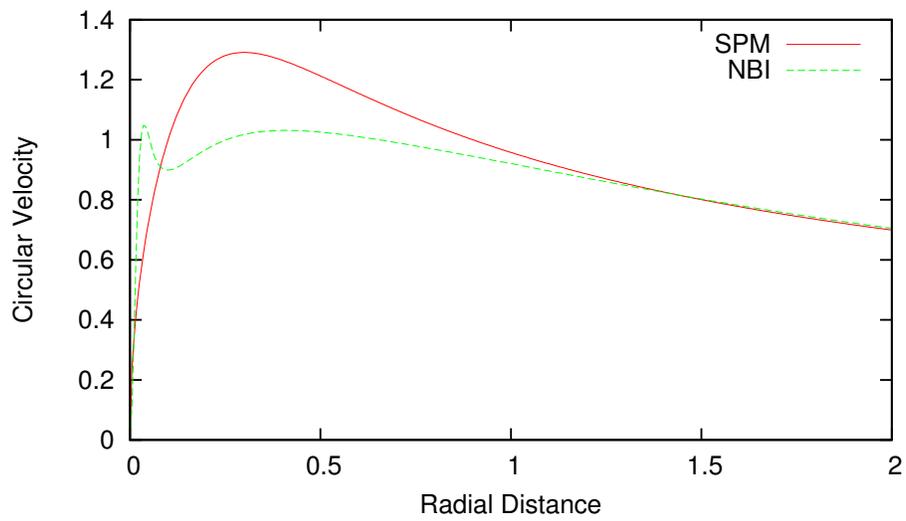}
\caption{The circular velocity as a function of radial distance for the two potentials.   The red solid SPM line designates the softened point mass potential described in equation \ref{spm-eqn} while the green dashed NBI line shows the potential for the n-body interaction potential described in equation \ref{nbi-eqn}.  The normalization for this graph is set so the asymptotic behavior is the same for both potentials.}
\label{circvel}
\end{center}
%\end{minipage}
\end{figure}

When studying interacting systems, it is the particles in the outer disk that are most likely to have their orbits perturbed.  The strength of the interaction is often described in terms of how closely the velocity of the secondary galaxy matches the orbital velocity of particles orbiting the primary.  For a given galaxy mass, the circular velocity of particles in the outer disk will be relatively lower in a softened point mass potential than what is observed in actual systems.  The new potential is an attempt to overcome that discrepancy.  The interpolated potential (which includes the effects of a dark matter halo) provides more realistic circular velocities for particles in the outer disk.  Figure \ref{circvel} shows the circular velocity as a function of radial distance from the center of the galaxy for the two potentials.  Each potential is calculated for the same total mass.  In the case of the n-body interpolated potential (NBI), the mass is distributed between the three components so the velocity is lower  in the disk compared to the softened point mass (SPM) potential.  Two important features to notice is that within the disk (r $\le$ 1.0) the NBI velocity is flatter than SPM and that outside the disk the two potentials give similar velocities.

\subsection{Dynamical Friction}
Self-consistent n-body codes have demonstrated that the orbits of secondary galaxies will decay over time.  One important process that leads to the loss of orbital energy is scattering in the form of dynamical friction.  These codes can also produce other multi-body effects like violent relaxation.  These effects are absent in restricted three-body codes.  The orbital decay, even during a first passage encounter, can be significant.  Dynamical friction plays a key role in galaxy evolution through other interactions such as between a bar and the dark matter halo.  A parameterized version of this effect leading to orbital decay was added to our code following the derivation in \cite{1987gady.book.....B}.

A massive body $M$ moving through a field of other massive particles will interact with them through the gravitational force.  The field particles have individual masses much less than $M$.  However, these field particles are part of an overall system that is very massive and large.  It is customary to approximate this system as infinite and homogeneous, with the distribution of velocities taken to be Maxwellian.  As the body $M$ moves through this field of stars, the field stars will be deflected resulting in an enhanced density behind the massive body sometimes referred to as a wake.  The attraction of this wake on the moving body is opposite in direction compared to its velocity resulting in dynamical friction.

For a set of background masses of density $\rho$ and a Maxwellian distribution of velocities with dispersion $\sigma$, Chandrasekhar's dynamical friction formula for the acceleration becomes
\begin{eqnarray}
\frac{d\boldsymbol{v}}{dt} & =&\frac{4 \pi G^2  M  {\rm ln}\Lambda }{v^3 } \left[ {\rm erf} \left( X \right) - \frac{  2 X }{ \sqrt{\pi}} e^ {- X ^2} \right]  \boldsymbol{v}
%\nonumber
\end{eqnarray}

\begin{eqnarray}
X = \frac{v}{\sigma}
\end{eqnarray}

We define the following values useful for calculating the velocity dispersion $\sigma$ as a fuction of $r$

\begin{eqnarray}
p(r) = G \int_r^{\infty} \frac {\rho(r)  m(r) }{r^2 } dr \nonumber \\
%\nonumber
\end{eqnarray}

\begin{eqnarray}
\sigma = v_r^2 = \frac{p(r)}{\rho(r)} \nonumber \\
%\nonumber
\end{eqnarray}

We calculate the density based upon the mass used in the three-component model.  In order to use the above formulas, this is equivalent to treating the bulge and disk masses as being distributed spherically rather than just in a disk.  This allows us to calculate the integrals by summing the density for a series of radially concentric shells around the galaxy.  Outside the truncation radius, we set $\rho(r)$ to zero and set $v_r$ equal to $p(r)$, which is just a velocity when divided by the ``unit'' mass, to avoid floating point problems.  The Coulomb logarithm, ln $\Lambda$, is set as a constant during the simulation.

\cite{petsch} describes a more sophisticated treatment of dynamical friction in a restricted three-body code.  Their approach includes the ability to account for density gradients that lead to a dynamical friction acceleration orthogonal to the direction of motion.  They also implemented four different models to calculate ln $\Lambda$.  Our approach is similar to the their ``Model A'', which is just a constant value.  Their ``Model B'' is distance-dependent, ``Model C'' is an interpolation between two constant values, and ``Model D'' is mass- and distance-dependent . \cite{petsch} perform comparisons between their restricted three-body simulations including dynamical friction with self-consistent n-body simulations.  They report reliable recreations of orbital decay for galaxy pairs with a mass ratio of three to one.  However, they conclude that this approach is not able to accurately recreate the decay for equal mass mergers.  The various effects felt by both galaxies through multiple close approaches are not likely to be modeled by simple analytic treatments.  This is one of the important reasons for performing full n-body simulations.  However, in cases where the two galaxies do not undergo multiple close approaches and are otherwise not simulated to final decay and merging, that the simplified dynamical friction treatment can still improve the accuracy of equal mass galaxy interactions.

We have introduced a number of approximations in our implementation of the dynamical friction force.  However, our implementation allows for rapid calculation of the force through simple interpolation.  This allows us to calculate the forces in our halo/disk/bulge model with dynamical friction in close to the same time as the softened point mass potential.  Any plausible treatment of dynamical friction is likely to yield more realistic results from restricted three-body simulations.  We believe our approximate implementation allows us to more easily match trajectories between our restricted three-body code and full n-body simulations.

\subsection{Coordinate System and Orbit Geometry}

\begin{figure}[htbp]
\begin{center}
\includegraphics{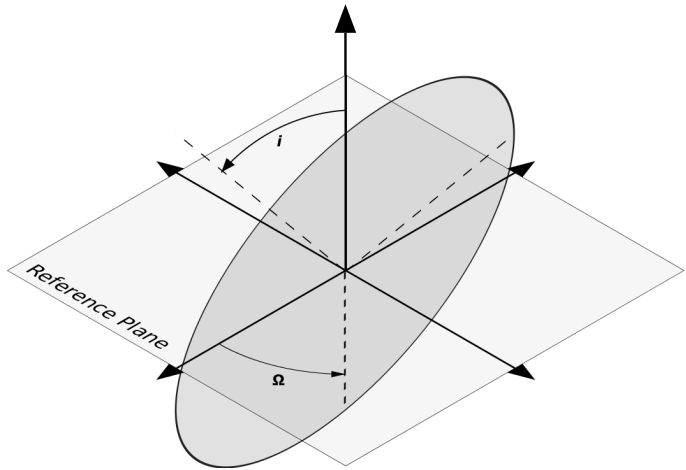}
\caption{The reference plane for the orbit geometry is the same as the plane of the primary disk for an unrotated galaxy.  Otherwise it is typical to treat the reference plane as the plane of the sky. }
\label{coords}
\end{center}
\end{figure}

The coordinate system of the orbit used in the model has its origin at the center of mass of the primary galaxy.  For modeling actual systems, the x-y plane coincides with the plane of the sky.  In these situations, the disk orientation angles of each galaxy, position angle and inclination, are used to rotate the respective disks to match the observed values.  To run the simulation in the plane of the disk of the primary galaxy, simply leave the orientation angles of the primary disk at zero.  Figure \ref{coords} shows a circle, representing the orbit plane of the secondary galaxy, inclined relative to the reference plane.  The inclination angle is marked as \emph{i} in the figure.  The line of nodes, where the orbit plane intersects the reference plane, is rotated in the reference plane by the argument of the ascending node, which is \emph{$\Omega$} in the figure.  To use the full range of Keplerian orbit parameters for initial conditions, it is recommended that the caller convert to position and velocity vectors for the relative orbit.

\subsection{Numerical Model}
The two galaxies to be simulated are initialized with particles placed randomly around the disk according to user-selected profiles and initialized with the appropriate circular velocities for the specified potentials.  The secondary galaxy, and its particles, is given a position and velocity relative to the primary galaxy.  After initialization in which the disks of each galaxy are rotated to the desired orientations, the model acceleration is calculated and used to integrate the position and velocity vectors for each particle in each time step.

\subsubsection{Initial Conditions}
\label{initc}

The particles for each galaxy are assigned an initial position and velocity by the PROFILE routine.  Particles are assigned to a radial distribution based on the user's choice of radial profile.  Particles can be distributed with a probability proportional to \( ^1/_r \), \( \exp(-r/r_{0}) \), or \( \exp(-r^{2}/a  +  r*b  +  c) \).  After determining a particle's radial position the circular velocity is calculated according to the user-specified central potential.  The particle's azimuthal position is then randomized.  Finally, an optional random ``heat'' value can be applied to the particle's velocity.  The user specifies the maximum magnitude of the velocity offset which is chosen from a uniform random distribution independently for each of the three components.

In order to simulate the changes to morphology experienced by interacting galaxies, the position and velocity of the secondary galaxy must be set at a time prior to the point of closest approach, where interaction effects are strongest.  However, for ease of comparison with other models, that often express the relative orbit of the interacting pair at the time of closest approach, we use backwards integration.  Users may specify the position and velocity of the secondary galaxy in two ways.  The first is by specifying several Keplerian orbit parameters at the time of closest approach.  The second is by providing the position and velocity vector of the secondary galaxy.  The position of the secondary galaxy and its test particles is set by the PERTURBER\_POSITION routine, which uses a backwards integration to calculate the position of the center of mass of the secondary galaxy according to the specified potential.  The integration works from time t=0 backwards to the user specified time.  The test particles associated with this galaxy have their positions and velocities updated based on the calculated starting values for the center of mass through a simple translation and then rotation to reflect the desired disk orientation.  In general, the backwards integration time is chosen so that the two galaxies are far enough apart that interaction effects are not yet causing the orbits of the test particles to evolve.

Assuming spherical symmetry the three-component potential is calculated for a set of radial bins according to the analytic potential described in Section \ref{threecomp}.  The contributions to the acceleration from each of the three component is summed into a single array. Arrays are also computed for mass, density, and velocity dispersion. Later, during subsequent force calculations, the radial distance between each particle and the respective centers-of-mass is used to lookup the force and velocity dispersion in this array.  The values are then scaled by multiplying by the appropriate mass and used in the force calculations.  It is this lookup process that gives the three-component potential its alternate name, of n-body interpolated potential (NBI).  

The period of interest for two interacting galaxies usually covers the time when they are relatively close to one another.  As such, the pre-computed arrays have a maximum application distance.  To increase the maximum distance covered by these arrays, callers should adjust the $rs\_internal$ parameter if the df\_module.f90 and NBIModel.java source files.  To maintain resolution of the pre-computed values, adjust the $nnn$ parameter in the same module.

\subsubsection{Integration}
Updated velocities and positions are calculated using a fixed time step, fourth-order Runge-Kutta integrator.  The force calculation for the selected potential is performed by the DIFFEQ routine.   This routine operates on an array of all particles, the two centers of mass and the test particles with each call.  

\section{Software Organization and Conventions}
The subroutine names, arguments, and types have been kept consistent between the FORTRAN and Java\textsuperscript{TM} implementations.  The included Makefile includes targets for compiling each of the language versions.

The integrator code remains the same regardless of potential used.  There is a separate implementation of the DIFFEQ routine for each potential: DIFFEQ\_SPM and DIFFEQ\_NBI.  Similarly the PROFILE routine calls the same CIRCULAR\_VELOCITY function to initialize particle velocities regardless of potential type.  Inside the CIRCULAR\_VELOCITY function, the calculation is different depending on whether the softened point mass or n-body interpolated potential is used.  The INIT\_DISTRIBUTION function is where the value for $ln \Lambda$ is set for dynamical friction.

\subsection{File Formats}

Particle positions and velocities are output in an ASCII text format.  The x-, y-, and z- positions and the x-, y-, and z- velocities are output as one line for each particle in fixed width columns of 16 characters each.  First the particles for the primary galaxy are output, followed by the particles for the secondary galaxy.  The position and velocity of the secondary galaxy are output as the last line in the file.

To view the particle positions simply plot the first and second columns of data as the x- and y- coordinates.  For convenience, the software can output a script file that can be used with the popular GNUPlot software package to view output from each of the time steps in an animated series.  The ASCII output file is also compatible for use with the Department of Energy VisIt\footnote{\url{https://wci.llnl.gov/simulation/computer-codes/visit}} tool.

The other major file used by JSPAM is a parameters file.  It is a text file of key-value pairs that allows control over the more than 30 simulation parameters.  The initialization routines have default values for these parameters.  Users have the choice of using a parameters file or modifying the initialization routines to set simulation parameters.  Additionally, the software can be passed a single command line argument containing the state information string from a simulation from the {\it Galaxy Zoo: Mergers} project.

\section{Results}
The precursor SPAM code has been used to produce a number of models of actual systems that have been published.  The enhanced JSPAM code has been used with {\it Galaxy Zoo: Mergers} project and has produced models for 62 systems of interacting galaxies \citep{holin}.

\subsection{Models of Specific Systems}
\begin{figure}[ht]
\begin{minipage}[b]{0.45\linewidth}
\centering
\includegraphics[width=\textwidth]{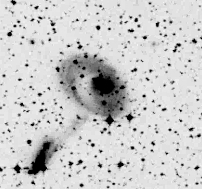}
\caption{DSS-provided UK Schmidt R-plate of AM 1724-622}
\label{mimg}
\end{minipage}
\hspace{0.5cm}
\begin{minipage}[b]{0.5\linewidth}
\centering
\includegraphics[width=\textwidth]{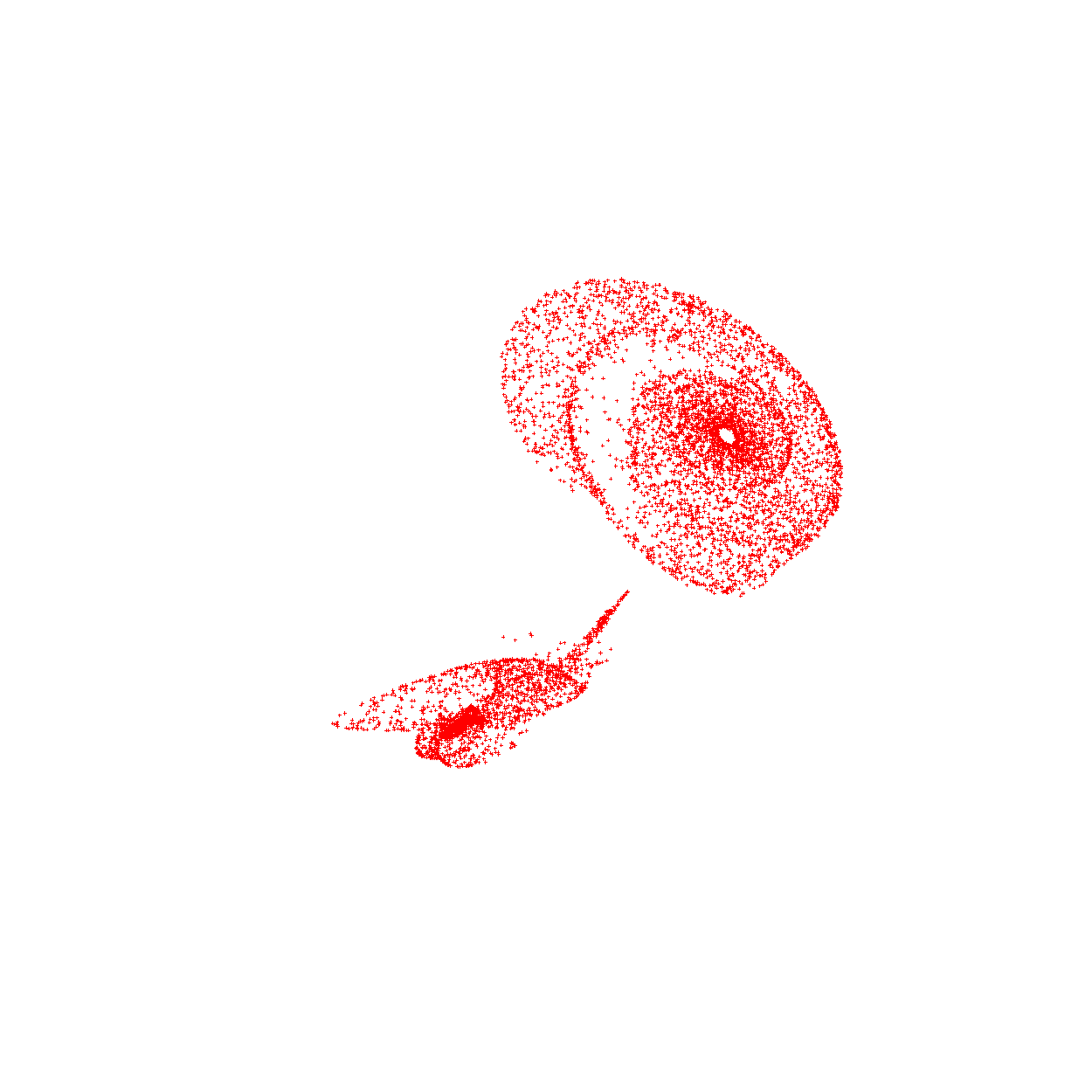}
\caption{JSPAM simulation of AM 1724-622 using the softened point mass potential.   }
\label{msim}

\end{minipage}
\end{figure}

Some of the systems modeled with the precursor SPAM code include the Sacred Mushroom \citep{wallin94} also known as AM 1724-622, NGC 2782 (Arp 215) \citep{smith94},  and NGC 7714/7715 (Arp 284) \citep{smith92}, \citep{struck03}.  For the Sacred Mushroom we present a DSS image of the system in Figure \ref{mimg} next to the simulation result in Figure \ref{msim}.  The general shape and density distortion of the primary disk are recreated for the mushroom head, and a tidal tail is formed that matches at least the orientation of the mushroom stalk.  The JSPAM code, which preserves the potential used in SPAM, was used to run the simulation shown in Figure \ref{msim}.

\begin{figure}[ht]
\begin{minipage}[b]{0.45\linewidth}
\centering
\includegraphics[width=\textwidth]{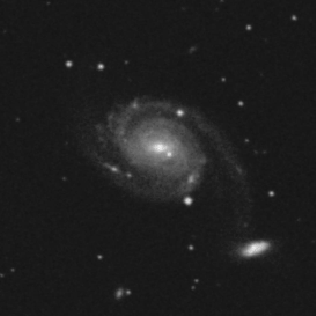}
\caption{SDSS image of Arp 86}
\label{a86img}
\end{minipage}
\hspace{0.5cm}
\begin{minipage}[b]{0.45\linewidth}
\centering
\includegraphics[width=\textwidth]{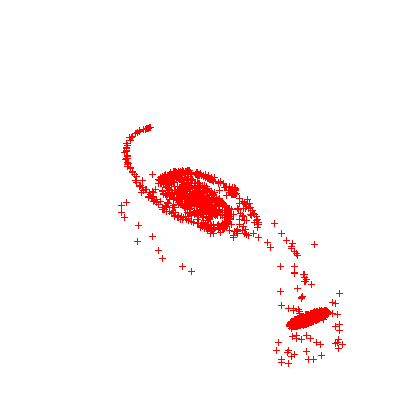}
\caption{JSPAM simulation of Arp 86}
\label{a86sim}
\end{minipage}
\end{figure}

The full details of the {\it Galaxy Zoo: Mergers} project will be published in a future paper\footnote{The {\it archive} directory of the source code at \url{https://github.com/jfwallin/JSPAM/} contains the input parameters and scripts necessary to recreate all 62 models discussed in \cite{holin}}.  Here we present a specific system, Arp 86, that was modeled during the development of that project \citep{wallin10}.  Figure \ref{a86img} shows an SDSS image of Arp 86 next to our simulation in Figure \ref{a86sim}.  The general position, orientation, and length of the tidal tails of the primary galaxy are recreated as well as the lack of significant distortion of the secondary galaxy.  The length of the northern tail on the primary may appear excessive compared to the grey scale thumbnail from SDSS, but matches the results of n-body simulations \citep{salo}.

\subsection{Timing Tests}

The simulation was run for 10 time units, approximately 870 Myr, with an integration timestep of 0.02 time units, 1.7 Myr.  The runtime for 1000 particles using the softened point mass potential on the test machine was about 300 ms excluding I/O.  This is the baseline runtime.  The average of 10 runs for several combinations of particle number and potential are reported as a ratio with the baseline runtime in Table \ref{timetbl}.  The runtime is linear with the number of particles up to 5000.  For 10000 and 20000, the storage required for particle state is greater 512KB.  This exceeds the L2 cache size per core on the test machine which resulted in super-linear increase in runtime.

\begin{table}
\caption[Timing Results]{The relative runtimes for the two potentials, softened point mass vs. n-body interpolated, for various numbers of particles. }
\begin{center}
  \begin{tabular}{ | l | c | r | }
    \hline
Particles & SPM & NBI \\ \hline
1,000 & 1.0 & 1.4 \\ \hline
2,000 & 2.0 & 2.7 \\ \hline
5,000 & 5.0 & 6.7 \\ \hline
10,000 & 12.0 & 15.0 \\ \hline
20,000 & 32.0 & 36.4 \\
    \hline
  \end{tabular}
\label{timetbl}
\end{center}
\end{table}

\subsection{Comparison to Full n-body Models}
\label{comparison}

The process of comparing the results of restricted three-body models with full n-body models begins with the backwards integration.  The secondary galaxy is moved backwards to the starting point.  The relative position and velocity of the secondary galaxy, along with disk orientation parameters, can then be transferred to the initial conditions for an n-body simulation.  The particles in the primary and secondary galaxies for the n-body simulation are initialized according to the mass distribution of the same three-component potential used in the restricted three-body code.  The particles in the secondary galaxy have their positions and velocities modified based on the results of the backwards integration, e.g. particles rotated according to disk orientation parameters, positions translated, and then center of mass velocity for the secondary galaxy added to each of its particles.  Advancing the n-body simulation forward to the same end time as the restricted three-body simulation should produce similar morphologies.  For many hyperbolic orbits where dynamical friction does not play an import role, the initial parameters from JSPAM  produce nearly identical simulations when they are compared to simulations run with codes such as GADGET \citep{gadget}, GADGET-2 \citep{gadget2}, and MASS99\footnote{MASS is the Multi-physics Astrophysical Simulation Software.} \citep{antunes_convergence_2001}.

\begin{figure}[htbp]
\begin{center}
\includegraphics[scale=0.7]{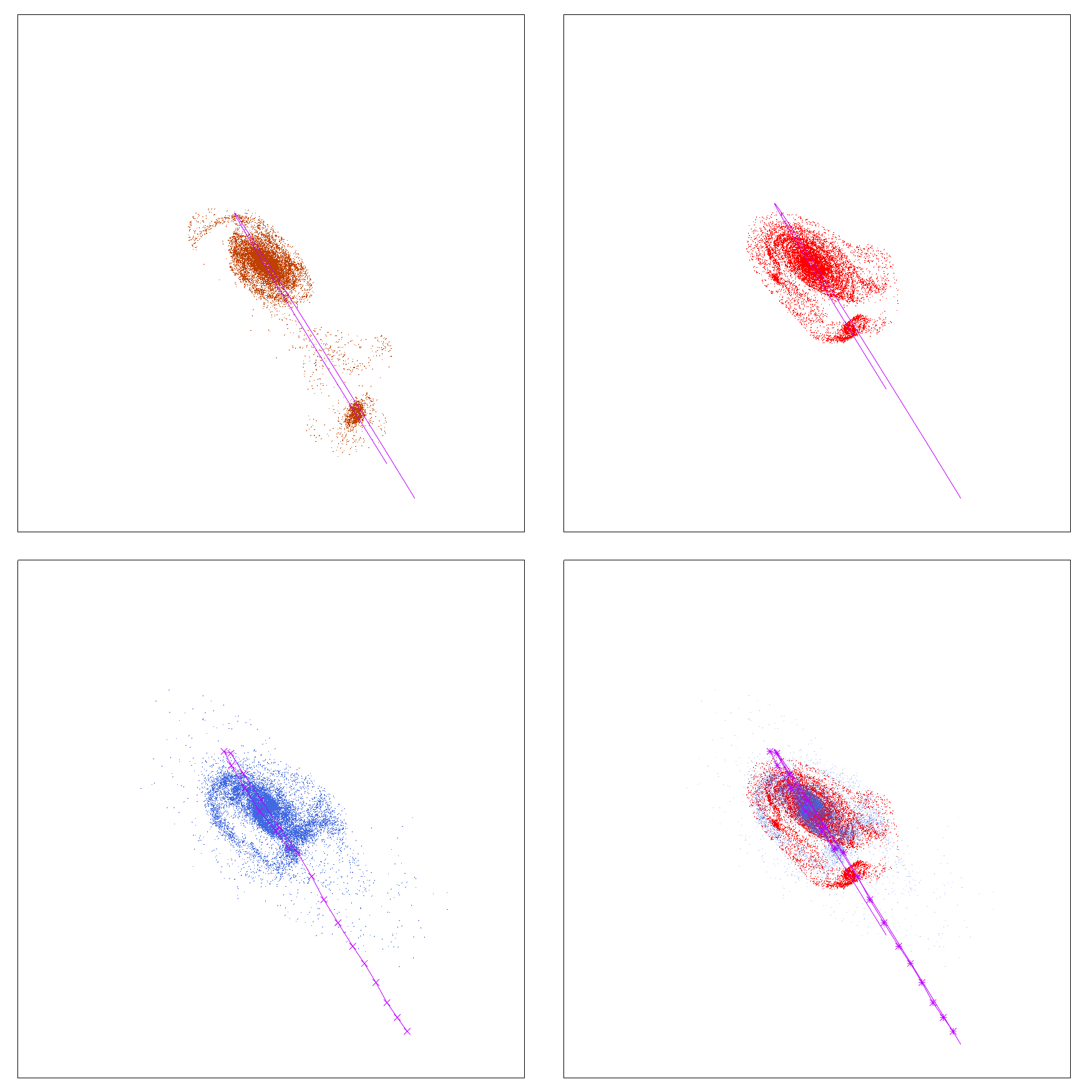}
\caption{Top left panel shows the final state of a simulation run using the softened point mass potential.  The top right panel shows the final state when run with the three-component potential of JSPAM.  The bottom left panel is a simulation run using the n-body tree code known as MASS99.  The bottom right panel shows the overlay with the three-component potential plotted on top of the MASS99 results.  All panels are of the x-z plane. The softened point mass in the top left panel is not a good match to the n-body results where as the three-component potential is.}
\label{a86nbody2}
\end{center}
\end{figure}

However, for closed orbits or slower parabolic orbits, dynamical friction will alter the trajectory of the orbits.  For the restricted three-body model the strength of dynamical friction is controlled by setting the value of Coulomb logarithm, ln $\Lambda$.  When performing the backwards integration step, the dynamical friction acts to increase the velocity of the secondary galaxy resulting in a different starting point.  Using this new starting point in a full n-body simulation will hopefully result in an end state that appears similar to the restricted three-body model.  By adjusting the value of  ln $\Lambda$, performing the restricted three-body backwards integration, and then the forwards n-body integration, it is possible to find similar end states.  This matching process was initially performed by hand.  \cite{harvey} has developed an automated process for adjusting ln $\Lambda$. The results for matching a restricted three-body simulation to an n-body simulation of Arp 86 run using MASS99 is show in Figure \ref{a86nbody2}.    

The JSPAM simulation shown in Figure \ref{a86nbody2} ran in about 15 seconds using the SPM potential and 16 seconds using the NBI potential.  On the same hardware, the full n-body tree code simulation takes several hours to complete at the resolution shown with 64000 particles integrated using MASS99 which has a similar run time to GADGET.  The SPM potential in the upper left panels of these figures is significantly different than the full n-body run shown in the bottom left panel.   That is different than results for the NBI potential which look very similar to n-body run.  Because self-gravity is not included, not all runs will necessarily be as similar as those shown in this figure.   However, this code can be effectively used to converge on orbital parameters for plausible matches to real galaxies while using significantly lower computational resources than self-consistent n-body codes.

\cite{holin} describes the results of the {\it Galaxy Zoo~: Mergers} project that modeled 62 pairs of observed interacting galaxies using the NBI potential with dynamical friction.  All 62 pairs were successfully modeled with comparable morphologies.  \cite{holin} also described several methods for estimating the uniqueness of each model as well as other convergence parameters.  The advantage of running an efficient code is that 10000 or more sets of initial conditions can be run for each system.  This gives better statistics for estimating uniqueness than only running a few dozen sets of initial conditions.  

The restricted three-body models from the {\it Galaxy Zoo~: Mergers} project were compared to n-body models by \cite{harvey}.  That work found that over 80\% of the systems were easily recreated with matching morphologies using n-body simulations.  Ultimately, researchers would like to study observed interacting systems with more sophisticated codes that include star formation and other physics.  We advocate using a multi-method approach that saves time and computational resources to narrow down most of the orbit parameters with a restricted three-body code.  By including a dynamical friction term our results can be used to initialize more complex n-body simulations with advanced physics.

\section{Conclusions}

The JSPAM code has produced restricted three-body simulations that successfully recreate the disturbed morphologies of actual interacting galaxies.  The enhanced potential based on a three-component model and including a treatment of dynamical friction further increases the ability to transition these simple models to full n-body simulations.   Because self-gravity is not included, this code will not produce all of the features in the more accurate n-body tree codes.  However, it can be utilized to effectively explore parameter space when modeling the interactions of observed interacting systems.  The restricted three-body code can be used to narrow down the orbit parameters and the more sophisticated codes can explore parameter space related to other effects of interactions such as star formation.  In particular, this code can be used with automatic algorithms to match galaxy simulations to real interacting systems via genetic algorithms or similar approaches. 

\section{License}

The JSPAM software is free software which can be redistributed and/or modified under the terms of the Academic Free License (``AFL'') v. 3.0.    The authors also request that academic works which make use of the software include a citation to this article.

\section{Acknowledgements}

The development of {\it Galaxy Zoo: Mergers} was supported by the US National Science Foundation under grant DRL-0941610.

Some images for this paper were taken from the Sloan Digital Sky Survey.
Funding for the SDSS and SDSS-II has been provided by the Alfred P. Sloan Foundation, the Participating Institutions, the National Science Foundation, the U.S. Department of Energy, the National Aeronautics and Space Administration, the Japanese Monbukagakusho, the Max Planck Society, and the Higher Education Funding Council for England. The SDSS Web Site is http://www.sdss.org/.

The SDSS is managed by the Astrophysical Research Consortium for the Participating Institutions. The Participating Institutions are the American Museum of Natural History, Astrophysical Institute Potsdam, University of Basel, University of Cambridge, Case Western Reserve University, University of Chicago, Drexel University, Fermilab, the Institute for Advanced Study, the Japan Participation Group, Johns Hopkins University, the Joint Institute for Nuclear Astrophysics, the Kavli Institute for Particle Astrophysics and Cosmology, the Korean Scientist Group, the Chinese Academy of Sciences (LAMOST), Los Alamos National Laboratory, the Max-Planck-Institute for Astronomy (MPIA), the Max-Planck-Institute for Astrophysics (MPA), New Mexico State University, Ohio State University, University of Pittsburgh, University of Portsmouth, Princeton University, the United States Naval Observatory, and the University of Washington.

%% in parentheses, after the keyword, but they will not be verified.

\bibliography{bibfile}

\end{document}